\documentclass{article}
\usepackage{amstex}
\newcommand{\bilin}[4]{\left(\frac{\delta}{\delta{#2}}{#3},
   {#1}\frac{\delta}{\delta{#2}}{#4}\right)}
\newcommand{\dilat}[1]{\left(D{#1},
   \frac{\delta}{\delta{#1}}\right)}
\newcommand{\dfour}[1]{\frac{{\mathrm d}^4{#1}}{(2\pi)^4}}
\newcommand{\dpartial}[1]{\frac{\partial}{\partial{#1}}}
\newcommand{\dscali}[1]{\frac{{\mathrm d}{#1}}{{#1}}}
\newcommand{\dtotal}[1]{\frac{{\mathrm d}}{{\mathrm d}{#1}}}

\newcommand{\laplace}[2]{\left(\frac{\delta}{\delta{#2}},
   {#1}\frac{\delta}{\delta{#2}}\right)}
\newcommand{\perm}[1]{{\mathfrak S}_{{#1}}}
\newcommand{\V}{\mathrm V}
\newcommand{\wt}{\widetilde}
\newcommand{\N}{{\mathbb N}}
\newcommand{\R}{{\mathbb R}}

\newenvironment{proof}{\noindent{\large Proof\/}:}{$\;\Box$}
\newenvironment{defin}{
   {\vspace{0.5\baselineskip}}\noindent{\large Definition\/}:}
   {\vspace{0.05\baselineskip}}
\newenvironment{hypot}{
   {\vspace{0.5\baselineskip}}\noindent{\large Hypothesis\/}:}
   {\par\vspace{0.05\baselineskip}}
\newtheorem{thm}{Theorem}[section]
\newtheorem{pro}[thm]{Proposition}
\newtheorem{lem}[thm]{Lemma}

\begin{document}

%\baselineskip+10mm

\begin{titlepage}
\begin{center}
{\Large \bf Running coupling expansion for \\[2mm]
the renormalized $\phi^4_4$--trajectory \\[2mm]
from renormalization invariance
}\\[10mm]
\end{center}
\begin{center}
{\Large C. Wieczerkowski} \\[10mm]
\end{center}
\begin{center}
Institut f\"ur Theoretische Physik I,
Universit\"at M\"unster, \\
Wilhelm-Klemm-Stra\ss e 9, D-48149 M\"unster, \\
wieczer@@uni-muenster.de \\[2mm]
\end{center}
\vspace{-10cm}
\hfill 
\parbox{4cm}{MS-TP1-96-18\\ hep-th/9612214}
\vspace{12cm}
\begin{abstract}

%\baselineskip+10mm

We formulate a renormalized running coupling expansion for
the $\beta$--function and the potential of the renormalized
$\phi^4$--trajectory on four dimensional Euclidean
space--time. Renormalization invariance is used as a first
principle. No reference is made to bare quantities. The
expansion is proved to be finite to all orders of perturbation
theory. The proof includes a large momentum bound on the
connected free propagator amputated vertices.
\end{abstract}
\end{titlepage}

\section{Introduction}
%=====================

We study the renormalization of a massless real scalar field
$\phi$ on four dimensional space--time, perturbed by a
$\phi^4$--vertex. Its renormalization is done by means of a
renormalization group transformation $R_L$, which scales by a
factor $L>1$. The renormalized theory comes as a pair,
consisting of a $\beta$--function $\beta(g)$ together with a
potential $V(\phi,g)$, both functions of the $\phi^4$--coupling
$g$ (but not of $L$), with the following properties:\\[2mm]
I) $V(\phi,g)$ is of the form
\begin{equation}
V(\phi,g)=g\int{\mathrm d}^4x\left\{\frac{\mu^{(1)}}{2}
\phi(x)^2+\frac{\zeta^{(1)}}{2}\phi(x)(-\bigtriangleup)
\phi(x)+\frac{1}{4!}\phi(x)^4\right\}+O(g^2).
\label{1.1}
\end{equation}
II) $V(\phi,g)$ is invariant in the sense that
\begin{equation}
(R_L V_{g_1})(\phi)=V_{g(L)}(\phi),
\label{1.2}
\end{equation}
where $g(L)$ is the solution to the flow equation
\begin{equation}
L\dtotal{L}g(L)=\beta(g(L))
\label{1.3}
\end{equation}
with the initial condition $g_1=g(1)$.\\[2mm]
We will show that there exists a $\beta$--function $\beta(g)$
and a potential $V(\phi,g)$, both unique to all orders of
perturbation theory in $g$, enjoying these properties (and
general qualities of a renormalized potential in $\phi^4$--theory).
The assignment $g\mapsto\{\beta(g),V(\phi,g)\}$ is called
the renormalized $\phi^4$--trajectory.

Renormalization invariance was introduced as a first principle in
\cite{Wi96}, both for a discrete and a continuous renormalization
group. The aim of the present paper is to give a short, in the sense 
of formal power series rigorous, construction of a renormalized
trajectory in perturbation expansion using the continuous 
renormalization group. Our inductive scheme follows Polchinski's proof 
of perturbative renormalizability \cite{P84}. Unlike Polchinski we will 
not start from a bare action, but directly compute the renormalized
theory. The setup will be close to that in \cite{Wi96}, with the 
difference that we will do without normal ordering here, and with the 
difference that we will admit a more general $\beta$-function. 
Indeed, the $\beta$--function can be brought to a normal form by means 
of a reparametrization of $g$ (which then becomes a function of the 
$\phi^{4}$--coupling). The normal form of our $\beta$-function in four 
dimensions is cubic. With a fixed $\beta$--function, the renormalized 
theory comes a fixed point of the renormalization group composed with 
a flow of the coupling, given by this $\beta$--function.  

\section{Renormalization group}

We begin on a formal level, which is strengthened as we move
towards perturbation theory.

Let $\phi$ be a real scalar field on four dimensional Euclidean
space--time. Consider the following renormalization group for
potentials $V(\phi)$, derived from a momentum space decomposition
of $(-\bigtriangleup)^{-1}$.

\begin{defin}
Let $R_L$ be the renormalization group transformation
\begin{equation}
(R_L\V)(\psi)=-\log\int{\rm d}\mu_{\Gamma_L}(\zeta)
\exp\left\{-\V(S_L\psi+\zeta)\right\}
+\mathrm{const.},
\label{2.1}
\end{equation}
depending on a scale parameter $L>1$, where
${\rm d}\mu_{\Gamma_L}(\zeta)$ denotes the Gaussian measure on
field space with mean zero and covariance
\begin{equation}
\widetilde{\Gamma_L}(p)=
\frac{\exp (-p^2)-\exp (-L^2\,p^2)}{p^2},
\label{2.2}
\end{equation}
and where $S_L$ denotes the the dilatation operator
\begin{equation}
\widetilde{S_L\psi}(p)=L^3\widetilde{\psi}(Lp).
\label{2.3}
\end{equation}
\end{defin}

Field independent constants are understood to be properly
removed. Notice that $\psi$ is rescaled with its canonical
scaling dimension. The renormalization group transformation
(\ref{2.1}) is a Gaussian convolution in rescaled form.

Concerning the background on the renormalization group,
we refer to Wilson and Kogut \cite{WK74}. A pedagocical
account of the perturbative momentum space renormalization
group can be found in the lectures by Benfatto and Gallavotti
\cite{BG95}. It was applied to the perturbative renormalization
of QED by Feldman, Hurd, Rosen, and Wright \cite{FHRW88}.

\begin{pro}
$R_L$ satisfies the semi--group property
\begin{equation}
R_L\circ R_{L^\prime}=R_{L\,L^\prime},\quad L,L^\prime >1,
\qquad \lim_{L\downarrow 1}R_L ={\mathrm id}.
\label{2.4}
\end{equation}
\end{pro}

Consequently, the iteration of (\ref{2.1}) with fixed $L$
is interpolated by an increase of $L$ in one transformation
(\ref{2.1}).

\begin{pro}
The renormalization group flow $\V(\psi,L)=(R_L\V)(\psi)$
satisfies the functional differential
equation
\begin{gather}
\left\{L\dpartial{L}-\dilat{\psi}\right\}\V(\psi,L)=
\nonumber\\
\laplace{C}{\psi}\V(\psi,L)
-\bilin{C}{\psi}{\V(\psi,L)}{\V(\psi,L)},
\label{2.5}
\end{gather}
where
\begin{equation}
\wt{D\psi}(p)=\left\{p\dpartial{p}+3\right\}\wt{\psi}(p),
\qquad
\wt{C}(p)=\exp(-p^2),
\label{2.6}
\end{equation}
with the initial condition $\V(\psi,1)=\V(\psi)$.
\end{pro}

The continuous renormalization group was invented by Wilson
\cite{WK74}. A review of its applications was given by
Wegner \cite{We76}. Its value in perturbative renormalization
was discovered by Polchinski \cite{P84}. Functional differential
equations for interpolated Gaussian convolutions are also
used in the cluster expansion of Glimm and Jaffe \cite{GJ87}.

An aim of renormalization theory is to construct renormalization
group flows which remain finite as $L\uparrow\infty$. A way
to proceed is to look for quantities which are independent of
$L$.

\begin{defin}
A scaling pair is a $\beta$--function $\beta(g)$ together with
a potential $V(\psi,g)$, both depending on a coupling $g$ but
not on $L$, such that
\begin{equation}
\V(\psi,L)=V(\psi,g(L))
\label{2.7}
\end{equation}
satisfies (\ref{2.5}) for any solution $g(L)$ of the ordinary
differential equation
\begin{equation}
L\dtotal{L}g(L)=\beta \left(g(L)\right).
\label{2.8}
\end{equation}
\end{defin}

A scaling potential is its own renormalization image in the
sense that
\begin{equation}
\left(R_L V_{g_1}\right)(\psi)=V_{g(L)}(\psi),
\label{2.9}
\end{equation}
where $g(L)$ is the solution of the one dimensional flow equation
(\ref{2.8}) to the initial condition $g(1)=g_1$. In view thereof,
$g(L)$ is called a running coupling.

\begin{pro}
A $\beta$--function $\beta(g)$ together with a potential $V(\psi,g)$
is a scaling pair if both together satisfy the functional differential
equation
\begin{gather}
\left\{\beta(g)\dpartial{g}-\dilat{\psi}\right\}V(\psi,g)=
\nonumber\\
\laplace{C}{\psi}V(\psi,g)-\bilin{C}{\psi}{V(\psi,g)}{V(\psi,g)}.
\label{2.10}
\end{gather}
\end{pro}

We will restrict our attention to Euclidean invariant
even potentials. Let $V(\psi,g)$ be given by a power series
\begin{gather}
V(\psi,g)=
\sum_{n=1}^{\infty}\frac{1}{(2n)!}\int\dfour{p_1}\ldots
\int\dfour{p_{2n}}(2\pi)^{4}\delta(\sum_{i=1}^{2n}p_{i})
\nonumber\\
\widetilde{\psi}(-p_1)\cdots\widetilde{\psi}(-p_{2n-1})
\widetilde{\psi}\left(\sum_{i=1}^{2n-1}p_i\right)
\;\wt{V}_{2n}(p_1,\ldots,p_{2n},g)
\label{2.11}
\end{gather}
in $\psi$. The question of its convergence will be left aside.
Let us instead identify $V(\psi,g)$ with its formal sequence of
vertices $\wt{V}_{2n}(p_1,\ldots,p_{2n},g)$. Vertices will be
restricted to the hyperplane of total zero momentum. They can
then be represented as
\begin{equation}
\wt{V}_{2n}(p_1,\ldots,p_{2n-1},-\sum_{i=1}^{2n-1}p_{i},g)=
\wt{V}_{2n}(p_{1},\ldots,p_{2n-1},g).
\label{2.11a}
\end{equation} 

\begin{pro}
A $\beta$--function $\beta(g)$ together with a potential
$V(\psi,g)$, viewed as a formal power series in $\psi$, is
a scaling pair if both together satisfy the system of
integro--differential equations
\begin{gather}
\left\{\beta(g)\dpartial{g}+\sum_{i=1}^{2n-1}p_i\dpartial{p_i}-
4+2n\right\}\wt{V}_{2n}(p_1,\ldots,p_{2n-1},g)=
\nonumber\\
\int\dfour{q}\;\wt{C}(q)\;
\wt{V}_{2(n+1)}(p_1,\ldots,p_{2n-1},q,-q,g)
\nonumber\\
-\sum_{m=1}^{n}\binom{2n}{2m-1}\Biggl[
\wt{C}\left(\sum_{i=1}^{2m-1}p_i\right)
\wt{V}_{2m}(p_1,\ldots,p_{2m-1},g)
\nonumber\\
\wt{V}_{2(n-m+1)}(p_{2m},\ldots,p_{2n-1},\sum_{i=1}^{2m-1}p_i,g)
\Biggr]_{\perm{2n-1}},
\label{2.12}
\end{gather}
where $[\cdot ]_{\perm{2n-1}}$ denotes the symmetrization in
$p_1,\ldots,p_{2n-1}$.
\end{pro}

The constant $4-2n$ is called the scaling dimension of a vertex.
Furthermore, vertices are called relevant, marginal, or irrelevant
when their scaling dimension is positive, zero, or negative.

\section{$\phi^4$--theory}

The non--irrelevant couplings of $V(\psi,g)$ play a
special role and deserve their own names. Let $\mu(g)$,
$\zeta(g)$, and $\lambda(g)$ be defined as
\begin{equation}
\mu(g)=\wt{V}_2(0,g),\quad
\zeta(g)=\dpartial{(p^2)}\wt{V}_2(p,g)\bigg\vert_{p=0},\quad
\lambda(g)=\wt{V}_4(0,0,0,g).
\label{3.1}
\end{equation}

A canonical choice of $g$ in $\phi^4$--theory is the value
of the quartic vertex at zero momentum. We prefer a slightly
more general definition.

\begin{defin}
Let $\lambda(g)$ be a given formal power series
\begin{equation}
\lambda(g)=
g+\sum_{r=2}^{\infty}\frac{g^r}{r!}\lambda^{(r)}
\label{3.2}
\end{equation}
in $g+g^2\R[[g]]$.
\end{defin}

The normalization $\lambda^{(1)}=1$ can always be achieved by
a rescaling of $g$. The choice $\lambda^{(r)}=0$, $r>1$, 
means selecting the $\phi^{4}$-coupling as expansion parameter.
Other choices serve to bring the $\beta$--function to a standard 
form. The cubic normal form will be discussed below. Any 
choice will do for the moment. $\lambda(g)$ will now be assumed
to be fixed.

We then expand both the $\beta$--function and the
vertices into power series in $g$,
\begin{gather}
\beta(g)=\sum_{r=1}^{\infty}\frac{g^r}{r!}\beta^{(r)},
\label{3.3}\\
\wt{V}_{2n}(p_1,\ldots,p_{2n-1},g)=\sum_{r=1}^{\infty}
\frac{g^r}{r!}\wt{V}_{2n}^{(r)}(p_1,\ldots,p_{2n-1}).
\label{3.4}
\end{gather}
Power series expansions for the couplings (\ref{3.1})
are included. All of them will be treated as formal power
series in $g$.

\begin{pro}
A $\beta$--function $\beta(g)$ together with a potential
$V(\psi,g)$, viewed as formal power series in $g$, is
a scaling pair if both together satisfy the system of
integro--differential equations
\begin{gather}
\left\{\sum_{i=1}^{2n-1}p_i\dpartial{p_i}-
4+2n+r\beta^{(1)}\right\}
\wt{V}_{2n}^{(r)}(p_1,\ldots,p_{2n-1})=
\nonumber\\
-\sum_{s=2}^{r}\binom{r}{s}\;\beta^{(s)}
\;\wt{V}_{2n}^{(r-s+1)}(p_1,\ldots,p_{2n-1})
\nonumber\\
+\int\dfour{q}\;\wt{C}(q)\;
\wt{V}_{2(n+1)}^{(r)}(p_1,\ldots,p_{2n-1},q,-q)
\nonumber\\
-\sum_{s=1}^{r-1}\binom{r}{s}\sum_{m=1}^{n}\binom{2n}{2m-1}
\Biggl[
\wt{C}\left(\sum_{i=1}^{2m-1}p_i\right)
\wt{V}_{2m}^{(s)}(p_1,\ldots,p_{2m-1})
\nonumber\\
\wt{V}_{2(n-m+1)}^{(r-s)}(p_{2m},\ldots,p_{2n-1},
\sum_{i=1}^{2m-1}p_i)
\Biggr]_{\perm{2n-1}}.
\label{3.5}
\end{gather}
\end{pro}

To obtain a mathematically well defined problem, we should say
what kind of solutions are looking for.

\begin{defin}
Let ${\mathbb V}$ be the space of vertices
$\wt{V}_{2n}^{(r)}(p_1,\ldots,p_{2n-1})$ with the following
properties:\\[2mm]
I) (Bose--symmetry)
\begin{equation}
\wt{V}_{2n}^{(r)}(p_{\pi(1)},\ldots,p_{\pi(2n)})=
\wt{V}_{2n}^{(r)}(p_{1},\ldots,p_{2n}),\qquad
\pi\in\perm{2n};
\label{3.6}
\end{equation}
II) ($O(4)$--symmetry)
\begin{equation}
\wt{V}_{2n}^{(r)}(R p_{1},\ldots,R p_{2n})=
\wt{V}_{2n}^{(r)}(p_{1},\ldots,p_{2n-1}),\qquad
R\in O(4);
\label{3.7}
\end{equation}
III) (Smoothness)
\begin{equation}
\wt{V}_{2n}^{(r)}\in
{\mathcal C}^{\infty}(\R^4\times\cdots\times\R^4);
\label{3.8}
\end{equation}
IV) (Large momentum bound)
\begin{gather}
\|\partial^\alpha\wt{V}_{2n}^{(r)}\|_{\infty,\epsilon}=
\nonumber\\
\sup_{(p_1,\ldots,p_{2n-1})\in\R^4\times\cdots\times\R^4}
\left\{\left\vert\partial^\alpha\wt{V}_{2n}^{(r)}(p_1,\ldots,p_{2n-1})
\right\vert
\exp\left(-\epsilon\sum_{i=1}^{2n-1}p_i^2\right)\right\}<\infty,
\nonumber\\
0<\epsilon < \frac{1}{2},\qquad
\alpha\in\N^4\times\cdots\times\N^4;
\label{3.9}
\end{gather}
V) (Connectedness)
\begin{equation}
\wt{V}_{2n}^{(r)}(p_1,\ldots,p_{2n-1})=0,\qquad
n>r+1;
\label{3.10}
\end{equation}
VI) (Coupling)
\begin{equation}
\wt{V}_4^{(r)}(0,0,0)=\lambda^{(r)},\qquad
r>1;
\label{3.11}
\end{equation}
VII) (Order one)
\begin{gather}
\wt{V}_2^{(1)}(p)=\mu^{(1)}+\zeta^{(1)}p^2,
\label{3.12}\\
\wt{V}_4^{(1)}(p_1,p_2,p_3)=1,
\label{3.13}\\
\wt{V}_{2n}^{(1)}(p_1,\ldots,p_{2n-1})=0,\qquad
n>2.
\label{3.14}
\end{gather}
\end{defin}

The properties (I),(II),(III), and (IV) are appropriate for
$\phi^{2N}$--theory, with any $N>1$, in this setup.
The properties (V), (VI), and (VII) distinguish $\phi^4$--theory.
See also Polchinski \cite{P84} and Keller, Kopper, and Salmhofer
\cite{KKS91}.\footnote{The authors use a cutoff function with
compact support. The large momentum bound is then unneccessary
as all loop integrals extend over a finite domain.}

\begin{thm}
(A) There exists a unique scaling pair in ${\mathbb V}$,
given by $\beta$--function $\beta(g)$ together with potential
$V(\psi,g)$, both viewed as formal power series in $g$,
whose vertices $\wt{V}_{2n}^{(r)}(p_1,\ldots,p_{2n-1})$
have the properties (I),...,(VII).
It is called the
renormalized $\phi^4$--trajectory.
(B) The $\beta$--function of the renormalized
$\phi^4$--trajectory is given by
$\beta(g)=\frac{-3}{(4\pi)^2}g^2+O(g^3)$.
The running coupling $g(L)$ is therefore asymptotically free
in the infrared direction.
\end{thm}

\noindent{\large Outline of the proof\/:}
The proof is an induction on $r$. The induction step
$r-1\rightarrow r$ consists of a sub--induction
$n+1\rightarrow n$, which goes backwards in the number of legs.
We compute $\wt{V}_{2n}^{(r)}(p_1,\ldots,p_{2n-1})$ in the
order $r+1,r,\ldots,1$. When coming to the case $n=2$, we
first compute $\beta^{(r)}$ and thereafter $\wt{V}^{(r)}_{4}
(p_1,p_2,p_3)$. In the case $n=1$, we first compute
the mass $\mu^{(r)}=\wt{V}_2^{(r)}(0)$, then
the wave function $\zeta^{(r-1)}=
\dpartial{(p^2)}\wt{V}_2^{(r-1)}(p)\vert_{p^2=0}$, and
thereafter $\wt{V}_2^{(r)}(p)$, except for $\zeta^{(r)}$.
Each of these steps will be shown to be both well defined
and to yield a unique solution.

\section{Proof of the Theorem}

To first order, (\ref{3.5}) simplifies to
\begin{gather}
\left\{\sum_{i=1}^{2n-1}p_i\dpartial{p_i}-4+2n+\beta^{(1)}\right\}
\wt{V}_{2n}^{(1)}(p_1,\ldots,p_{2n-1})=
\nonumber\\
\int\dfour{q}\wt{C}(q)\;
\wt{V}_{2(n+1)}^{(1)}(p_1,\ldots,p_{2n-1},q,-q).
\label{4.1}
\end{gather}

\begin{lem}
The first order vertices, given by (\ref{3.12}), (\ref{3.13}),
and (\ref{3.14}), satisfy (\ref{4.1}) if and only if
\begin{equation}
\beta^{(1)}=0,\qquad
\mu^{(1)}=\frac{-1}{2(4\pi)^2}.
\label{4.2}
\end{equation}
The first order coupling $\zeta^{(1)}$ is a free
parameter.
\end{lem}

\begin{proof}
For $n=2$, (\ref{3.13}) and (\ref{3.14}) satisfy (\ref{4.1})
if $\beta^{(1)}=0$. For $n=1$, (\ref{3.12}) and (\ref{3.13})
satisfy (\ref{4.1}) if
\begin{equation}
\mu^{(1)}=\frac{-1}{2}\int\dfour{q}\wt{C}(q).
\label{4.3}
\end{equation}
This integral is convergent and evaluated to (\ref{4.2}).
\end{proof}

$\mu^{(1)}$ is a normal ordering constant for the first
order quartic vertex. $\zeta^{(1)}$ is better thought of
as a second order quantity. Its value will be computed
from a second order equation.

\begin{hypot}
Suppose that we have determined all coefficients $\beta^{(s)}$
and all vertices $\wt{V}_{2m}^{(s)}(p_1,\ldots,p_{2m-1})$, for
$1\leq s\leq r-1$ and $1\leq m\leq s+1$, except for the
coupling $\zeta^{(r-1)}$. Suppose further that we have determined
$\wt{V}_{2m}^{(r)}(p_1,\ldots,p_{2m-1})$, for $n+1\leq m\leq
r+1$. Suppose that all vertices, determined so far, have the
properties (I),...,(VII). We proceed with the computation of
$\wt{V}_{2n}^{(r)}(p_1,\ldots,p_{2n-1})$ under these assumptions.
\end{hypot}

To save space we write
\begin{gather}
\wt{K}_{2n}^{(r)}(p_1,\ldots,p_{2n-1})=
-\sum_{s=2}^{r}\binom{r}{s}\;\beta^{(s)}
\;\wt{V}_{2n}^{(r-s+1)}(p_1,\ldots,p_{2n-1})
\nonumber\\
+\int\dfour{q}\;\wt{C}(q)
\;\wt{V}_{2(n+1)}^{(r)}(p_1,\ldots,p_{2n-1},q,-q)
\nonumber\\
-\sum_{s=1}^{r-1}\binom{r}{s}\sum_{m=1}^{n}\binom{2n}{2m-1}
\Biggl[
\wt{C}\left(\sum_{i=1}^{2m-1}p_i\right)
\wt{V}_{2m}^{(s)}(p_1,\ldots,p_{2m-1})
\nonumber\\
\wt{V}_{2(n-m+1)}^{(r-s)}\left(p_{2m},\ldots,p_{2n-1},
\sum_{i=1}^{2m-1}p_i\right)
\Biggr]_{\perm{2n-1}},
\label{4.4}
\end{gather}
for the right hand side of (\ref{3.5}).

\begin{lem}
The integral in (\ref{4.4}) is convergent.
The differential vertex, given by (\ref{4.4}), has the
properties (\ref{3.6}), (\ref{3.7}), (\ref{3.8}), (\ref{3.9}),
and (\ref{3.10}).
\end{lem}

\begin{proof}
We prove the large momentum bound (\ref{3.9}) for the case
of no momentum derivatives. Use part of the expontential decay
of $\wt{C}(q)$ for an $L_{\infty,\epsilon}$--bound
on $\wt{V}_{2(n+1)}^{(r)}(p_1,\ldots,p_{2n-1},q,-q)$.
Put an $L_1$--bound on the remaining one loop integral.
The result is an estimate
\begin{gather}
\|\wt{K}_{2n}^{(r)}\|_{\infty,\epsilon}\leq
\sum_{s=2}^{r-1}\binom{r}{s}\vert\beta^{(s)}\vert
\;\|\wt{V}_{2n}^{r-s+1}\|_{\infty,\epsilon}
+\frac{C_0}{(1-2\epsilon)^2}
\;\|\wt{V}_{2(n+1)}^{(r)}\|_{\infty,\epsilon}
\nonumber\\
+\sum_{s=1}^{r-1}\binom{r}{s}\sum_{m=1}^{n}\binom{2n}{2m-1}
\;\|\wt{V}_{2m}^{(s)}\|_{\infty,\epsilon}
\;\|\wt{V}_{2(n-m+1)}^{(r-s)}\|_{\infty,\epsilon},
\label{4.5}
\end{gather}
where $C_0$ is a constant, independent of $r$ and $n$.
Momentum derivatives are distributed on all factors,
which are then estimated along the same lines.
The other assertions are elementary.
\end{proof}

Therefore, we have a well defined first order partial differential
equation
\begin{equation}
\left\{\sum_{i=1}^{2n-1}p_i\dpartial{p_i}-4+2n\right\}
\wt{V}_{2n}^{(r)}(p_1,\ldots,p_{2n-1})=
\wt{K}_{2n}^{(r)}(p_1,\ldots,p_{2n-1})
\label{4.6}
\end{equation}
for the vertex labelled by $n$ and $r$. The perturbative scaling
dimension is $4-2n$, independent of $r$, since $\beta^{(1)}=0$.
The induction is put in such an order that the right hand side
of (\ref{4.6}) is known from the previous work. It is directly
integrated in the irrelevant case $4-2n <0$.

\begin{lem}
For $n> 2$, (\ref{4.6}) has a unique solution with the properties
(\ref{3.6}), (\ref{3.7}), (\ref{3.8}), and (\ref{3.9}). It is
given by the convergent integral
\begin{equation}
\wt{V}_{2n}^{(r)}(p_1,\ldots,p_{2n-1})=
\int_{0}^{1}\dscali{L}L^{-4+2n}
\wt{K}_{2n}^{(r)}(Lp_1,\ldots,Lp_{2n-1}).
\label{4.7}
\end{equation}
\end{lem}

\begin{proof}
Equation (\ref{4.6}) is equivalent to
\begin{equation}
L\dtotal{L}\left\{L^{-4+2n}\;
\wt{V}_{2n}^{(r)}(Lp_1,\ldots,Lp_{2n-1})
\right\}=
L^{-4+2n}\;
\wt{K}_{2n}^{(r)}(Lp_1,\ldots,Lp_{2n-1}),
\label{4.8}
\end{equation}
which is integrated to (\ref{4.7}). The difference of two
solutions to (\ref{4.8}) satisfies the homogeneity condition
\begin{equation}
L\dtotal{L}\left\{L^{-4+2n}\;
\bigtriangleup\wt{V}_{2n}^{(r)}(Lp_1,\ldots,Lp_{2n-1})
\right\}=0.
\label{4.9}
\end{equation}
Regularity at $p_1=\cdots =p_{2n-1}=0$ excludes solutions
thereof other than zero. Therefore, (\ref{4.7}) is unique.
The other properties are obvious.
\end{proof}

The irrelevant part of the potential has thereby been
determined. (\ref{4.7}) is a recursion relation for the
irrelevant vertices. Notice that (\ref{4.7}) evaluates to
zero in the case $n>r+1$. Therefore, the property
(\ref{3.10}) iterates to the next order.

\begin{lem}
For $n>2$, the vertices given by the integral (\ref{4.7}),
are independent of $\zeta^{(r-1)}$.
\end{lem}

\begin{proof}
The differential vertex (\ref{4.4}) is a linear function of
$\zeta^{(r-1)}$ with
\begin{gather}
\dpartial{\zeta^{(r-1)}}
\wt{K}_{2n}^{(r)}(p_1,\ldots,p_{2n-1})=
-\binom{r}{2}\;\beta^{(2)}\;p_1^2\;\delta_{n,1}
\nonumber\\
-2\binom{r}{1}\binom{2}{1}
\left(\mu^{(1)}+\zeta^{(1)}p_1^2\right)\;p_1^2\;\delta_{n,1}
\nonumber\\
-2\binom{r}{1}\binom{4}{1}\wt{C}(p_1+p_2+p_3)
\; (p_1+p_2+p_3)^2\;\delta_{n,2},
\label{4.10}
\end{gather}
zero for $n\geq 3$. The assertion follows by induction on $n$.
\end{proof}

We come to the non--irrelevant cases.
We cannot integrate the differential equations (\ref{3.5})
for the quadratic and the quartic vertex directly to
(\ref{4.7}). The  non--negative scaling dimension, $4-2n\geq0$,
causes a divergence at $L=0$. This problem is cured
by a Taylor expansion with remainder.

Consider first the quartic vertex. The differential equation
(\ref{4.8}) for the quartic vertex is
\begin{equation}
L\dtotal{L}\wt{V}_4^{(r)}(Lp_1,Lp_2,Lp_3)=
\wt{K}_4^{(r)}(Lp_1,Lp_2,Lp_3).
\label{4.11}
\end{equation}
Let us separate the coupling $\lambda^{(r)}$ from the
quartic vertex according to
\begin{equation}
\wt{V}_4^{(r)}(p_1,p_2,p_3)=
\lambda^{(r)}+\int_{0}^{1}{\mathrm d}L
\frac{{\mathrm d}}{{\mathrm d}L}
\wt{V}_4^{(r)}(Lp_1,Lp_2,Lp_3).
\label{4.12}
\end{equation}
Recall that we have fixed $\lambda^{(r)}$ by definition of $g$.
Evaluate (\ref{4.11}) at $L=0$, to conclude that the differential
quartic kernel has to vanish at zero momentum. This condition
determines $\beta^{(r)}$. The Taylor remainder can be computed
from
\begin{equation}
\left\{L\dtotal{L}+1\right\}\dtotal{L}
\wt{V}_4^{(r)}(Lp_1,Lp_2,Lp_3)=
\dtotal{L}\wt{K}_4^{(r)}(Lp_1,Lp_2,Lp_3),
\label{4.13}
\end{equation}
obtained by taking one $L$--derivative of (\ref{4.11}). The
gain of one $L$--derivative is thus one unit of scaling dimension,
whereupon we are back in the irrelevant case.

\begin{lem}
The differential equation (\ref{4.11}) has smooth solutions
only if
\begin{equation}
\wt{K}_{4}^{(r)}(0,0,0)=0.
\label{4.14}
\end{equation}
This condition is fulfilled if and only if
\begin{gather}
\beta^{(r)}=
-\sum_{s=2}^{r-1}\binom{r}{s}\;\beta^{(s)}\;\lambda^{(r-s+1)}
\nonumber\\
+\int\dfour{q}\wt{C}(q)\wt{V}_{6}^{(r)}(0,0,0,q,-q)
-\sum_{s=1}^{r-1}\binom{r}{s}\;2\binom{4}{1}
\mu^{(s)}\;\lambda^{(r-s)},
\label{4.15}
\end{gather}
where $\lambda^{(1)}=1$.
\end{lem}

I cannot resist from computing the second order coefficient
$\beta^{(2)}$ at this instant. The six--point--vertex in
second order is computed to\footnote{Consequently,
$\wt{V}_6^{(2)}(0,0,0,q,-q)=-8-12\frac{1-e^{-q^2}}{q^2}$.}
\begin{equation}
\wt{V}_6^{(2)}(p_1,\ldots,p_5)=-20
\left[\frac{1-e^{-(p_1+p_2+p_3)^2}}{(p_1+p_2+p_3)^2}
\right]_{\perm{5}}
\label{4.16}
\end{equation}
by means of (\ref{4.7}), with $r=2$ and $n=3$.
The integral in (\ref{4.15}) of it is elementary. Using the value
(\ref{4.2}) of $\mu^{(1)}$ it follows that
\begin{equation}
\beta^{(2)}=\frac{-6}{(4\pi)^2},
\label{4.17}
\end{equation}
which is the expected result. The negative sign has an
important consequence. It tells that the flow on the
renormalized $\phi^4$--trajectory at weak coupling is
asymptotically free in the infrared direction.

\begin{lem}
(A) The differential equation (\ref{4.13}) has a unique solution
with the properties (\ref{3.6}), (\ref{3.7}), (\ref{3.8}),
and (\ref{3.9}). It is given by the convergent integral
\begin{equation}
L\dtotal{L}\wt{V}_4^{(r)}(Lp_1,Lp_2,Lp_3)=
\int_0^L\dscali{L^\prime}L^\prime
\dtotal{L^\prime}\wt{K}_4^{(r)}
(L^\prime p_1,L^\prime p_2,L^\prime p_3).
\label{4.18}
\end{equation}
(B) The differential equation (\ref{4.11}) has a unique solution
with the properties (\ref{3.6}), (\ref{3.7}), (\ref{3.8}),
and (\ref{3.9}). It is given by the convergent integral
\begin{equation}
\wt{V}_4^{(r)}(p_1,p_2,p_3)=
\lambda^{(r)}
+\int_0^1\dscali{L} \wt{K}_4^{(r)}(Lp_1,Lp_2,Lp_3).
\label{4.19}
\end{equation}
\end{lem}

\begin{proof}
The proof of (A) is the same as that of (\ref{4.7}).
One $L$--derivative is just enough to fall into the case of
negative scaling dimension. Concerning (B), we notice that the
integral (\ref{4.19}) converges because
\begin{equation}
\wt{K}_4^{(r)}(Lp_1,Lp_2,Lp_3)=O(L)
\label{4.20}
\end{equation}
for all $(p_1,p_2,p_3)\in\R^4\times\R^4\times\R^4$ due to
the condition (\ref{4.14}).
The large momentum bound on (\ref{4.19}) follows from the estimate
\begin{equation}
\|\wt{V}_4^{(r)}\|_{\infty,\epsilon}\leq
\vert\lambda^{(r)}\vert
+\frac{C_1}{\epsilon}\sum_{i=1}^3\sum_{\mu=1}^4
\left\|\dpartial{p_i^\mu}\wt{V}_4^{(r)}\right\|_{\infty,\epsilon},
\label{4.21}
\end{equation}
where $C_1$ is a constant which is independent of $r$. The
large momentum bound on the momentum derivatives of the
quartic vertex follows from similar estimates. The other
assertions are obvious.
\end{proof}

Notice that the large momentum bound is not uniform in
$\epsilon$. This is the price for the Taylor expansion.
We then come to the quadratic vertex. Its personal
differential equation reads
\begin{equation}
\left\{L\dtotal{L}-2\right\}\wt{V}_2^{(r)}(Lp)=
\wt{K}_2^{(r)}(Lp).
\label{4.22}
\end{equation}
We represent it by a Taylor formula of order two with
remainder. Because of Euclidean invariance, we have that
\begin{equation}
\wt{V}_{2}^{(r)}(p)=
\mu^{(r)}+\zeta^{(r)}\;p^2+
\frac{1}{2}\int_{0}^{1}{\mathrm d}L (1-L)^2
\frac{{\mathrm d}^3}{{\mathrm d}L^3}
\wt{V}_{2}^{(r)}(Lp).
\label{4.23}
\end{equation}
We follow a similar procedure as in the case of the quartic vertex.
The Taylor remainder is computed as solution to the differential
equation
\begin{equation}
\left\{L\dtotal{L}+1\right\}
\frac{{\mathrm d}^3}{{\mathrm d}L^3}\wt{V}_2^{(r)}(Lp)=
\frac{{\mathrm d}^3}{{\mathrm d}L^3}\wt{K}_2^{(r)}(Lp).
\label{4.24}
\end{equation}
Three $L$--derivatives have brought us back to the irrelevant
case.

\begin{lem}
The differential equation (\ref{4.22}) has smooth solutions
(\ref{4.23}) only if
\begin{equation}
-2\mu^{(r)}=\wt{K}_{2}^{(r)}(0),
\qquad
0=\dpartial{(p^2)}\wt{K}_2^{(r)}(p)\biggr\vert_{p^2=0}.
\label{4.25}
\end{equation}
These conditions are fulfilled if and only if
\begin{gather}
\mu^{(r)}=\frac{1}{2}\biggl\{
\sum_{s=2}^{r}\binom{r}{s}\beta^{(s)}
\mu^{(r-s+1)}
\nonumber\\
-\int\dfour{q}\;\wt{C}(q)\;\wt{V}_4^{(r)}(0,q,-q)
+2\sum_{s=1}^{r-1}\binom{r}{s}
\mu^{(s)}\mu^{(r-s)}\biggr\}.
\label{4.26}
\end{gather}
and
\begin{gather}
\zeta^{(r-1)}=
\frac{-1}{\binom{r}{2}\beta^{(2)}+
2\binom{r}{1}\binom{2}{1} \mu^{(1)}}\times
\nonumber\\
\biggl\{\sum_{s=3}^{r}\binom{r}{s}\beta^{(s)}
\zeta^{(r-s+1)}
-\dpartial{(p^2)}\int\dfour{q}\; \wt{C}(q)\;
\wt{V}_4^{(r)}(p,q,-q)\biggr\vert_{p^2=0}
\nonumber\\
+\sum_{s=1}^{r-1}\binom{r}{s}\binom{2}{1}\mu^{(s)}
\mu^{(r-s)}
+\sum_{s=1}^{r-2}\binom{r}{s}\binom{2}{1}2
\zeta^{(s)}\mu^{(r-s+1)}\biggr\}.
\label{4.27}
\end{gather}
The order $r$ wave function $\zeta^{(r)}$ is a free parameter.
\end{lem}

Notice that both $\mu^{(r)}$ and $\zeta^{(r-1)}$ are finite
numbers. The integrals in (\ref{4.26}) and (\ref{4.27}) are
convergent. The denominator of the first factor on the RHS
of eq. (\ref{4.27}) is different from zero. 
Notice further that $\mu^{(r)}$, as given by
(\ref{4.26}), is independent of $\zeta^{(r-1)}$. The remaining
work is easily put to order.

\begin{lem}
(A) The differential equation (\ref{4.24}) has a unique
integral with the properties
(\ref{3.7}), (\ref{3.8}), and (\ref{3.9}).
It is given by
\begin{equation}
\frac{{\mathrm d}^3}{{\mathrm d}L^3}
\wt{V}_2^{(r)}(Lp)=
\int_0^L\dscali{L^\prime}L^\prime
\frac{{\mathrm d}^3}{{\mathrm d}{L^\prime}^3}
\wt{K}_2^{(r)}(L^\prime p).
\label{4.28}
\end{equation}
(B) The quadratic vertex, assembled through (\ref{4.23}),
is unique up to the parameter $\zeta^{(r)}$. For any finite
value of $\zeta^{(r)}$, it satisfies the properties
(\ref{3.7}), (\ref{3.8}), and (\ref{3.9}).
\end{lem}

\begin{proof}
(A) is another application of the integral (\ref{4.7}).
(B) is put together from (A). The large momentum bound on
the quadratic kernel follows from
\begin{equation}
\|\wt{V}_2^{(r)}\|_{\infty,\epsilon}\leq
\vert\mu^{(r)}\vert+\frac{C_2}{\epsilon}\vert\zeta^{(r)}\vert
+\frac{C_3}{\epsilon^2}\;\left\|\frac{\partial^2}{\partial (p^2)^2}
\wt{V}_2^{(r)}\right\|_{\infty,\epsilon}
\label{4.29}
\end{equation}
with some constants $C_2$ and $C_3$, both independent of $r$.
Similar estimates hold for all momentum derivatives.
\end{proof}

The quadratic remainder depends on $\zeta^{(r-1)}$
but not on $\zeta^{(r)}$. The estimate for the quadratic kernel
is valid for any finite value of $\zeta^{(r)}$.\\[3mm]
\noindent{\large Induction\/:}
We have shown that all assumptions of the induction hypothesis
are valid to order $r$ if they are valid up to order $r-1$.
Since they are fulfilled to order one, they iterate to all
orders of perturbation theory. The proof is complete.

\section{$\beta$-function}

The $\beta$-function transforms under reparametrizations as 
a vector field. Consider reparametrizations of formal power 
series. It follows that $\beta^{(2)}$ and $\beta^{(3)}$ are
universal, i.e., are not changed under reparametrizations. 
The other coefficients are not universal. We have showed that
a $\beta$-function with finite coefficients exists for all
choices of $\lambda (g)$. It is straight forward to determine
the reparametrisation inductively order by order which brings
all higher coefficients $\beta^{(r)}$, $r>3$, to zero. This 
is a canonical $\beta$-function for the renormalization 
group as a dynamical system. There is a direct implementation
of this idea. Instead of imposing a condition on $\lambda (g)$
at the beginning we could have imposed a condition on 
$\beta (g)$, saying that it should be exactly cubic. 
Recall that $\lambda^{(1)}$ was normalized to one. It turns
out that $\lambda^{(2)}$ can always be reparametrized to 
zero for the cubic $\beta$-function. Eq. (\ref{4.15}) is now
used as follows.The second order equation determines 
$\beta^{(2)}$, the third order equation determines 
$\beta^{(3)}$, and the order $r+1$-equation, $r>3$, determines
$\lambda^{(r)}$. Again all coefficients follow from convergent
integrals and are hence finite.

\section{Conclusions}

The identification of renormalized trajectories as invariant manifolds 
of renormalization group fixed points is part of Wilson's explanation 
of universality \cite{WK74}. It opens a way to study renormalized 
field theories without recourse to bare quantities. The essence of our 
renormalization group problem is to associate with a pair, given by a 
fixed point and an eigenvector of the linearized transformation (at this 
fixed point), an invariant curve, at the fixed point tangent to the 
eigenvector, together with a $\beta$--function. This scheme applies 
to unstable manifolds, center manifolds, and stable manifolds; with
merely different $\beta$--functions. This method was introduced in 
\cite{Wi96} using a momentum space renormalization group. See also the
previous studies \cite{WX95,RW96} of local potential approximations.
See \cite{M96} for an application to the massive renormalized 
trajectory of the scalar three dimensional infrared fixed point in 
(derivative expansion about) a local potential approximation. In the 
language of dynamical systems, we are here computing an invariant curve 
in the center manifold of the trivial fixed point, whose tangent at the 
trivial fixed point is a (normal ordered) $\phi^4$--vertex. The 
potential of the technology from the theory of dynamical systems in 
renormalization theory has been pointed out by Eckmann and Wittwer
\cite{EW84}, and by Gawedzki, Kupiainen, and Tirozzi \cite{GKT85}.

The renormalization of Euclidean quantum fields to all orders of
perturbation theory has been streamlined considerably by means of
the renormalization group. We mention the work of Callan
\cite{C76} (using the field theoretic renormalization group),
Gallavotti \cite{G85}, and Polchinski \cite{P84} (using Wilson's
renormalization group \cite{WK74}). We also mention the subsequent
contributions of Lesniewski \cite{L83}, Gallavotti and Nicol\`o
\cite{GN85}, Hurd \cite{H89}, Keller, Kopper, and Salmhofer
\cite{KKS91}. The recursion relation, furnished by (\ref{4.7}), 
(\ref{4.15}), (\ref{4.18}), (\ref{4.25}), (\ref{4.26}), and 
(\ref{4.27}), is the most direct renormalization scheme known to me.

The renormalization group transformation (\ref{2.1}) is a
Gaussian convolution in rescaled form. An intrinsic scale is
missing. When applied to a description of elementary particles,
this scheme requires an additional datum: a renormalization
scale. We have used a dimensionless formalism where all
quantities are expressed in units of this renormalization scale.

It remains to be seen whether renormalization invariance is a 
solid starting point for non--perturbative studies. 
The non--perturbative construction of the renormalized 
$\phi^{4}$--trajectory in the local potential approximation 
below four dimensions has been recently accomplished in 
\cite{Wi97}.

\end{document}